\documentclass[12pt]{iopart}

\usepackage{bm}
\usepackage{epsfig}
\usepackage{citesort}
\usepackage{graphicx}
\eqnobysec

\newcommand{\lwig}{\mbox{\;\raisebox{.3ex}
    {$<$}$\!\!\!\!\!$\raisebox{-.9ex}{$\sim$}\;}}
\newcommand{\gwig}{\mbox{\;\raisebox{.3ex}
    {$>$}$\!\!\!\!\!$\raisebox{-.9ex}{$\sim$}}\;}
\newcommand{\lambdabar}%
{{\hbox{$\lambda$\kern-1.ex\raise+0.45ex\hbox{--}}}}

\long\def\dump#1{}

\begin{document}


\begin{flushright}
{\large \tt MPP-2007-50 \\
TUM-HEP-668/07}
\end{flushright}

\title{Observational bounds on the cosmic radiation density}

\author{J.~Hamann$^1$, S.~Hannestad$^2$,
G.~G.~Raffelt$^3$ and Y.~Y.~Y.~Wong$^3$}

\address{$^1$~Physik Department T30e,
 Technische Universit\"at M\"unchen\\
 James-Franck-Strasse, D-85748 Garching, Germany\\
 $^2$~Department of Physics and Astronomy\\
 University of Aarhus, DK-8000 Aarhus C, Denmark\\
 $^3$~Max-Planck-Institut f\"ur Physik (Werner-Heisenberg-Institut)\\
 F\"ohringer Ring 6, D-80805 M\"unchen, Germany}

\ead{\mailto{jan.hamann@ph.tum.de},
     \mailto{sth@phys.au.dk},
     \mailto{raffelt@mppmu.mpg.de} and \\
     \mailto{ywong@mppmu.mpg.de}}

\begin{abstract}
We consider the inference of the cosmic radiation density,
traditionally parameterised as the effective number of neutrino
species $N_{\rm eff}$, from precision cosmological data.  Paying
particular attention to systematic effects, notably scale-dependent
biasing in the galaxy power spectrum, we find no evidence for a
significant deviation of $N_{\rm eff}$ from the standard value of
$N_{\rm eff}^0=3.046$ in any combination of cosmological data sets,
in contrast to some recent conclusions of other authors. The
combination of all available data in the linear regime prefers, in
the context of a ``vanilla+$N_{\rm eff}$'' cosmological model,
$1.1<N_{\rm eff}<4.8$ (95\%~C.L.) with a best-fit value of~2.6.
Adding data at smaller scales, notably the Lyman-$\alpha$ forest,
we find $2.2<N_{\rm eff}<5.8$ (95\%~C.L.) with~3.8 as the best fit.
Inclusion of the Lyman-$\alpha$ data shifts the preferred
$N_{\rm eff}$ upwards because the $\sigma_8$ value derived from the SDSS
Lyman-$\alpha$ data is inconsistent with that inferred from CMB. In
an extended cosmological model that includes a nonzero mass for
$N_{\rm eff}$ neutrino flavours, a running scalar spectral index and
a $w$ parameter for the dark energy, we find $0.8<N_{\rm eff}<6.1$
(95\%~C.L.) with 3.0 as the best fit.
\end{abstract}
\maketitle

\section{Introduction}                        \label{sec:introduction}

The observed global properties of the universe can be remarkably well
described by the $\Lambda$CDM model in conjunction with simple initial
conditions for the primordial density fluctuation spectrum.  In its
simplest form the model is geometrically flat and
represented by nontrivial values for six key parameters: the baryon
density, the dark matter density, the Hubble parameter, the amplitude
and spectral index of primordial adiabatic scalar fluctuations, and
the optical depth to reionisation. No single additional parameter
provides a substantially better fit to currently available data,
a situation
summarised by Max Tegmark's dictum,
``vanilla rules ok'' \cite{Tegmark:2006az}.

There are however many ways to extend this vanilla model,
some of which are physically well-motivated, such as a nontrivial
equation of state $p=w\rho$ for the dark energy, or a running
spectral index for the spectrum of primordial density fluctuations.
An extension with a nonvanishing hot dark matter component is
actually unavoidable because neutrinos are known to have mass and
the current direct laboratory limits are so loose that neutrino hot
dark matter could easily play an important role. Many authors have
sought to constrain neutrino masses in the context of $\Lambda$CDM
cosmology by inference from cosmological data, and found no evidence
for a nonvanishing value on the level of precision that can be
achieved with existing data.

Another extension invokes a nonstandard radiation density,
traditionally pa\-ram\-eterised by the effective number $N_{\rm eff}$
of neutrino species, with $N_{\rm eff}^0=3.046$ being the standard value 
\cite{nnu}.
This tradition dates back to the time before LEP at CERN measured
the number of ordinary neutrino species to be~3 and big bang
nucleosynthesis (BBN) provided the only significant upper limit on the
number of particle families. Today, constraining $N_{\rm eff}$ with
cosmological data is primarily a consistency test of standard particle
physics with concordance cosmology and of concordance cosmology with
itself because one can compare the radiation density allowed by BBN
with that implied by precision cosmological data which probe physics
at different epochs.

This exercise has been performed by several groups before
\cite{bib:hannestad2003,Crotty:2003th,Pierpaoli:2003kw,Barger:2003zg,Crotty:2004gm,%
Hannestad:2003ye} and
after~\cite{Seljak:2006bg,Hannestad:2006mi,%
Cirelli:2006kt,Ichikawa:2006vm,Mangano:2006ur,Friedland:2007vv,Spergel:2006hy} the
release of the WMAP 3-year data
\cite{Spergel:2006hy,Hinshaw:2006ia,Page:2006hz}. Some of these
recent results suggest surprisingly large values for $N_{\rm eff}$,
with 95\% C.L.~intervals that do not always include the standard
value $N_{\rm eff}^0=3.046$
\cite{Seljak:2006bg,Mangano:2006ur,Spergel:2006hy}. The apparent
conflict of these results and the exciting possibility
of a deviation from the minimal cosmology has motivated us to
re-examine the cosmological $N_{\rm eff}$ determination. Our goals
are two-fold: first, to identify the source of discrepancy in
previous analyses, and second, to provide an up-to-date estimation
of $N_{\rm eff}$ within more general model frameworks.

One possible source for the overestimation of $N_{\rm eff}$ is an
incorrect statistical methodology. The popular software {\sc
GetDist}, an analysis package frequently used in conjunction with
the Monte Carlo Markov Chain generator {\sc CosmoMC}
\cite{Lewis:2002ah,cosmomc} for cosmological parameter estimation,
provides by default 1D error estimates based on the central rather
than the minimal credible interval, although the latter is more
meaningful for inference problems. These constructions differ
significantly for skewed distributions, but become identical in the
Gaussian limit. We find that this effect can indeed be significant
if one uses a small number of data sets that are not very
constraining, since in these cases the 1D marginal posterior
distribution for $N_{\rm eff}$ often has a long tail towards large
$N_{\rm eff}$ values as a result of strong degeneracies with other
parameters. However, when many data sets are combined and conspire
to remove these degeneracies, the 1D posterior for $N_{\rm eff}$
usually becomes narrow enough to approach the Gaussian limit.
Therefore the different error construction methods are probably not
the main source of discrepancy.

The two main problems we have identified that affect the
determination of $N_{\rm eff}$ are (i) an unusually large
fluctuation amplitude reconstructed from the Lyman-$\alpha$ forest
data \cite{mcdonald} relative to that inferred from WMAP, and (ii)
the treatment of scale-dependent biasing in the galaxy power
spectrum inferred from the main galaxy sample of the Sloan Digital
Sky Survey data release 2 (SDSS-DR2)
\cite{Tegmark:2003uf,Tegmark:2003ud}. The first issue is well known,
and its complete investigation---involving elaborate astrophysical
modelling---is beyond the scope of the present work. The second
issue is more subtle. In previous analyses, scale-dependent biasing
in SDSS-DR2 has either been ignored \cite{Spergel:2006hy}, or
treated with empirical correction formulae under overly restrictive
conditions \cite{Seljak:2006bg,Mangano:2006ur}. We will explain this
issue in more detail in section~\ref{sec:bias} below. Here we
anticipate that no exotic values for $N_{\rm eff}$ will be found if
one either avoids small-scale data altogether or if one avoids
artificially constraining assumptions about the extent of the scale
dependence.

To derive our estimate for $N_{\rm eff}$ we begin in
section~\ref{sec:models} with a description of our cosmological
parameter framework, and in section~\ref{sec:data} the cosmological
data to be used. In section~\ref{sec:bias} we discuss the problem of
galaxy bias and its scale dependence. In section~\ref{sec:inference}
we compare different statistical inference methods frequently
encountered in the context of cosmological parameter estimation, and
the way they provide ``best-fit parameters'' and associated error
estimates. In section~\ref{sec:minimal} we study $N_{\rm eff}$ in a
minimal cosmological model which has a nonstandard radiation density
as the only extension to  vanilla cosmology. We use this simple
scenario as a benchmark to compare results from different
combinations of data and with different statistical methods. In
section~\ref{sec:extended} we consider an extended model that
includes as free parameters also a constant dark energy equation of
state parameter, a running spectral index, and neutrino masses. In
the framework of standard Bayesian statistics we provide credible
intervals for $N_{\rm eff}$. In section~\ref{sec:conclusions} we
summarise our findings.

\section{Cosmological models}                       \label{sec:models}

We perform our inference in the framework of a cosmological model
with vanishing spatial curvature and described by eleven
free parameters,
\begin{equation}\label{eq:model}
{\bm \theta} = \{\underbrace{\omega_{\rm
dm},\omega_b,H_0,\tau,\ln(10^{10}A_s),n_s}_{\rm vanilla},
 f_\nu,N_m,N_{\rm eff},w,\alpha_s\}.
\end{equation}
Here, the physical dark matter density $\omega_{\rm dm}=\Omega_{\rm
dm} h^2$, the baryon density $\omega_b=\Omega_b h^2$, the Hubble
parameter $H_0=h~100~{\rm km~s^{-1}~Mpc^{-1}}$, the optical depth to
reionisation $\tau$, the amplitude $A_s$, and the spectral index
$n_s$ of the primordial scalar power spectrum are collectively
labelled the ``vanilla'' parameters. They represent the simplest parameter
set necessary for a consistent interpretation of currently
available data.

The next three parameters denote a nonzero neutrino fraction $f_\nu
= \Omega_\nu/\Omega_{\rm dm}$ of the present day dark matter
content, the number $N_m$ of massive neutrino species, assuming a
common mass value $m_\nu$ for all of them, and the total effective
number $N_{\rm eff}$ of massless plus massive neutrinos. Of course,
$N_{\rm eff}$ can also include other forms of radiation. With
these definitions, $N_m$ enters the present-day energy density as
\begin{equation}
 \Omega_\nu h^2 =
 \frac{N_m m_\nu}{93 \, {\rm eV}}
 = \frac{\sum m_\nu}{93 \, {\rm eV}}.
\end{equation}
During the radiation-domination epoch the total energy density is
\begin{equation}
 \rho = \frac{\pi^2}{30}\,T_\gamma^4
 \left[2 + 2 \times \frac{7}{8} N_{\rm eff}
 \left(\frac{T_\nu}{T_\gamma}\right)^4 \right],
\end{equation}
where $T_\gamma$ and $T_\nu$ are the photon and neutrino
temperatures respectively.

The last two parameters in equation~(\ref{eq:model}) represent a
constant equation of state parameter for the dark energy $w$, and a
running parameter $\alpha_s$ in the scalar power spectrum defined
at the pivot scale $k=0.002~{\rm Mpc}^{-1}$.

The vanilla cosmological model is defined by holding all non-vanilla
parameters fixed at their standard values given in
table~\ref{tab:priors}.  In the same table we also show the priors
assumed for all cosmological fit parameters. We shall consider
several  scenarios, each including $N_{\rm eff}$ as a free
parameter.

\begin{table}
\caption{Standard values and priors for our cosmological fit
parameters. Prior~2 is identical to prior~1 except for the Hubble
parameter. All priors are uniform in the given intervals (i.e., top
hat). Depending on the investigated scenario, we use either the
standard value, or one of the priors for each
parameter.\label{tab:priors}} \hskip25mm
{\footnotesize
\begin{tabular}{llll}
 \br
 Parameter&Standard&Prior~1&Prior~2\\
 \mr
 $\omega_{\rm dm}$  &---& $0.01$--$0.99$ \\
 $\omega_{\rm b}$   &---& $0.005$--$0.1$ \\
 $h$                &---& $0.2$--$2.0$ &$ 0.4$--$1.0$\\
 $\tau$             &---& $0.01$--$0.8$ \\
 $\ln(10^{10}A_s)$ &---& $2.7$--$4.0$ \\
 $n_s$              &---& $0.5$--$1.5$ \\
 $f_\nu$            &0 & $0$--$0.5$ \\
 $N_m$              &0&$0$--$50$ \\
 $N_{\rm eff}$   &3.046&$0$--$50$\\
 $w$             &$-1$&$-2$--$0$\\
 $\alpha_s$         &0&$-0.2$--$0.2$\\
 \br
\end{tabular}
}
\end{table}


\subsection*{Minimal model}

Our minimal model (section~\ref{sec:minimal}) has seven free
parameters, namely, vanilla+$N_{\rm eff}$, while the other
parameters are fixed at their standard values. In particular, all
neutrinos are assumed to be massless. Most constraints on $N_{\rm
eff}$ in the recent literature were derived within this framework
\cite{Spergel:2006hy,Seljak:2006bg,Cirelli:2006kt,%
Ichikawa:2006vm,Mangano:2006ur}. Therefore, the minimal model lends
itself as a benchmark case to the study of differences and
similarities between our results and those of previous authors, as
well as differences between different analysis methods.

\subsection*{Extended models}

As in the minimal model, our extended models
(section~\ref{sec:extended}) always include the vanilla parameters
and $N_{\rm eff}$. In addition, we include neutrino masses and hence
the parameter $f_\nu$. Extended models with $f_\nu$ as a free
parameter were also considered in Refs.~\cite{bib:hannestad2003,%
Crotty:2004gm,Hannestad:2003ye,Hannestad:2006mi}.
However, there are many different ways to incorporate neutrino
masses into the analysis.  We shall consider two scenarios.  In the
first, we assume that all degrees of freedom represented by $N_{\rm
eff}$ have equal mass $m_\nu$, i.e.,  $N_m=N_{\rm eff}$. An
increased effective number density of ordinary neutrinos could be
due to, for example, a chemical potential in the neutrino phase
space.%
\footnote{Technically, even though a chemical potential 
does increase the neutrino number density, our treatment does not fully 
cover this case because it entails a neutrino velocity dispersion 
different from the standard non-degenerate Fermi--Dirac distribution.}

A second way to include neutrino masses, to be denoted $^3f_\nu$, is
to fix $N_m=N_{\rm eff}^0=3.046$, i.e., the standard density of
ordinary neutrinos, each with a mass $m_\nu$, is guaranteed. The
remaining $N_{\rm eff}-N_{\rm eff}^0$ species are massless degrees
of freedom that truly represent radiation; we do not assume anything
about its physical nature. The prior $N_{\rm eff}^0<N_{\rm eff}<50$
will be used in this case.

In both cases we consider also more elaborate scenarios in which $w$
and $\alpha_s$ are treated as free parameters, motivated by the
well-known degeneracies between $N_{\rm eff}$ and $f_\nu$
\cite{bib:hannestad2003}, and between $N_{\rm eff}$ and $w$
\cite{Hannestad:2005gj}. Studying these larger models and comparing
them with simpler ones illustrates how well combinations of
different data sets can break these degeneracies.

\section{Data}                                        \label{sec:data}

\subsection{Cosmic microwave background (CMB)}

We use CMB data from the Wilkinson Microwave Anisotropy Probe (WMAP)
ex\-per\-i\-ment after three years of observation
\cite{Spergel:2006hy,Hinshaw:2006ia,Page:2006hz}. The data analysis
is performed using version 2 of the likelihood calculation package
provided by the WMAP team on the LAMBDA homepage \cite{lambda}.

\subsection{Large scale structure (LSS)}

The large scale matter power spectrum has been inferred from the
galaxy clustering data of the Sloan Digital Sky Survey (SDSS)
\cite{Tegmark:2006az,Percival:2006gt,Tegmark:2003uf,Tegmark:2003ud}
and the Two-degree Field Galaxy Redshift Survey
(2dF)~\cite{Cole:2005sx}. In particular, the luminous red galaxies
(LRG) sample from the recent SDSS data release~5 (DR5) supersedes
all previous power spectrum measurements in terms of statistical
significance~\cite{Tegmark:2006az,Percival:2006gt}. However, the
``old'' spectrum retrieved from the SDSS main galaxy sample from
data release~2 (SDSS-DR2) \cite{Tegmark:2003uf,Tegmark:2003ud} is
still drawing attention, primarily because the parameter estimates
inferred therefrom appear to be in conflict with those derived from
other probes. We shall therefore analyse this data set as well.  As
it turns out, the apparent discrepancy can be explained in terms of
scale-dependent bias (section~\ref{sec:bias}).

\subsection{Baryon acoustic oscillations (BAO)}

The baryon acoustic oscillations peak has been measured in the SDSS
luminous red galaxy sample \cite{Eisenstein2005}.  We use all
20~points in the two-point correlation data set supplied in
Ref.~\cite{Eisenstein2005} and the analysis procedure described
therein, including power spectrum dewiggling, nonlinear corrections
with the {\sc Halofit} package~\cite{halofit}, corrections for
redshift-space distortion, and analytic marginalisation over the
normalisation of the correlation function.  
Except for the last marginalisation, these corrections are 
applied largely for cosmetic reasons; we obtain essentially the same results 
even without them.

\subsection{Type Ia supernovae (SNIa)}

We use the luminosity distance measurements of distant type~Ia
supernovae provided by Davis et~al.~\cite{Davis:2007na}. This sample
is a compilation of supernovae measured by the Supernova Legacy
Survey (SNLS) \cite{Astier:2005qq}, the ESSENCE project
\cite{Wood-Vasey:2007jb}, and the Hubble Space Telescope
\cite{Riess:2006fw}, as well as a set of 45 nearby supernovae. In
total the sample contains 192 supernovae.

\subsection{Hubble space telescope key project (HST)}

In some cases we use the direct measurement of the Hubble
parameter from the HST key project, $H_0=72
\pm 8 \ {\rm km \ s}^{-1} \ {\rm Mpc}^{-1}$~\cite{Freedman:2000cf}.

\subsection{Lyman-$\alpha$ forest (Ly$\alpha$)}
\label{sec:lya}

Measurements of the flux power spectrum of the Lyman-$\alpha$ forest
has been used to reconstruct the matter power spectrum on small scales
at large redshifts. By far the largest sample of spectra comes from
the SDSS survey. This data set was carefully analysed in
McDonald et~al.~\cite{mcdonald} and used to constrain the linear matter
power spectrum. The derived linear fluctuation amplitude at
$k=0.009~{\rm km~s}^{-1}$ and $z=3$ is
$\Delta^2 = 0.452^{+0.07}_{-0.06}$, and the effective spectral index
 $n_{\rm eff} = -2.321^{+0.06}_{-0.05}$. These results were
derived using a very elaborate model of the local intergalactic
medium in conjunction with hydrodynamic simulations.

While the Ly$\alpha$ data provides in principle a very powerful
probe of the fluctuation amplitude on small scales, the question
remains as to the level of systematic uncertainty in the result. The
same data has been reanalysed by Seljak et~al.\ \cite{Seljak:2006bg}
and Viel et~al.\ \cite{Viel:2005eg,Viel:2005ha,viel2006}, with
somewhat different results. Specifically, the normalisation found in
Refs.~\cite{Viel:2005eg,Viel:2005ha,viel2006} is lower than that
reported in Ref.~\cite{mcdonald}.

We shall use the default Ly$\alpha$ module provided in the {\sc CosmoMC}
package in some parts of our analysis.
This module uses the SDSS-Ly$\alpha$ data based on
McDonald et~al.~\cite{mcdonald}, and does not support the 
parameters $f_\nu$, $w$ and $\alpha_s$
in our extended models (it does support $N_{\rm eff}$, however).  Therefore, the Ly$\alpha$ data
will be analysed only in the context of the minimal model.

We stress that our Ly$\alpha$ results would likely be somewhat different if
the Viel et~al.\ analysis of SDSS-Ly$\alpha$
had been used.  However, when all available cosmological data sets are
used in combination, the Ly$\alpha$ data
carries relatively little weight in the combined fit for $N_{\rm eff}$ and
is not crucial for our conclusions.

\section{Scale-dependent bias}                        \label{sec:bias}

The conventional wisdom behind using galaxy survey data to infer the
underlying matter distribution is that, on sufficiently large
scales, the galaxy power spectrum $P_{\rm g}$ traces that of the
total matter content $P_{\rm m}$ calculated from linear theory up to
a constant, scale-independent bias factor,
\begin{equation}\label{eq:constantbias}
 P_{\rm g}(k)=b^2 P^{\rm lin}_{\rm m}(k).
\end{equation}
This relation is of course not exact, and its region of
applicability limited. On sufficiently small scales we expect
nonlinear evolution to cause its breakdown.

One obvious source for correction is the nonlinear growth of the
underlying matter density field on scales $k \gwig k_{\rm nl} \sim
0.15 \ h \  {\rm Mpc}^{-1}$. Another is the violation of
scale independence for the galaxy bias. The latter arises from the
fact that galaxy formation takes place preferentially in dark matter
halos with certain optimal masses, which are themselves biased
tracers of the matter
distribution~\cite{Benson:1999mv,Blanton:1999gd}. Indeed, depending
on the galaxy morphology, theoretical modelling and numerical
simulations suggest that the galaxy bias can deviate markedly from
scale independence already at nominally linear scales $k  \sim  0.1\
h \  {\rm Mpc}^{-1}$ \cite{Seljak:2000jg,Smith:2006ne}. The problem
this presents to cosmological parameter estimation is immediate:
power spectrum measurements on scales in the vicinity of $k  \sim
0.1\ h \  {\rm Mpc}^{-1}$ carry substantial weight in statistical
inferences because of their small formal error bars. Improper
handling of the galaxy bias will therefore likely yield misleading
results, a point we discuss in more detail below.

Unfortunately, neither theoretical modelling nor simulations are as
yet able to accurately predict the galaxy bias and its scale
dependence. In the meantime, we have the option to either (i)~cut
the data at a suitably small $k_{\rm max}$, usually $k_{\rm max}
\lwig 0.1 \ h \ {\rm Mpc}^{-1}$, or, if we want to use more data
points, (ii)~introduce some fitting formula that models crudely the
effect of a scale-dependent bias and then marginalise over the
associated nuisance parameters. For the latter approach and in the
framework of $\Lambda$CDM cosmologies, Ref.~\cite{Cole:2005sx}
suggests the formula
\begin{equation}\label{eq:fittingformula}
 P_{\rm g}(k)=b^2\,\frac{1+Q_{\rm nl} k^2}{1+A_g k}\,
 P^{\rm lin}_{\rm m}(k)\,,
\end{equation}
where $A_g=1.4$ is fixed, and $b$ and $Q_{\rm nl}$ are free
parameters to be marginalised. While the issue of bias correction
was not explored in the parameter estimation analysis of
SDSS-DR2~\cite{Tegmark:2003ud}, both options (i) and (ii) were
considered in the context of the vanilla model by the
2dF~\cite{Cole:2005sx} and the SDSS-DR5~\cite{Tegmark:2006az} teams
in their respective analyses. Both analyses found that, after
marginalisation over $Q_{\rm nl}$, additional data beyond $k \sim
0.1 \ h \ {\rm Mpc}^{-1}$ in option~(ii) lead to no significant
deviation in the cosmological parameter estimates or improvement in
the errors compared to those obtained with the simpler option~(i).

Conversely, if we ignore the issue of scale-dependent bias and adhere
strictly to the relation (\ref{eq:constantbias}), then it has been
shown that the 2dF-inferred $\Omega_m$ tends towards higher values
with increasing $k_{\rm max}$ \cite{Cole:2005sx}.  More strikingly,
analyses of the SDSS-DR5 data
show that the best-fit $\Omega_m$ values
inferred on scales $0.01 < k/(h \ {\rm Mpc}^{-1}) < 0.06 $ and $0.01 <
k/(h \ {\rm Mpc}^{-1}) < 0.15$ differ by 2--$3\sigma$ under the constant
bias assumption~(\ref{eq:constantbias})~\cite{Percival:2006gt}. Significant
scale dependence in the galaxy bias has been put forward to explain
the apparent tension between the galaxy power spectra measured by 2dF
and SDSS,
the latter of which tends to select the more strongly-biased red galaxies
\cite{Percival:2006gt,Cole:2006kn}. For the purpose of constraining a
possible nonstandard radiation density, we note that the well-known
degeneracy between $N_{\rm eff}$ and $\Omega_m$ means that any
inference of $N_{\rm eff}$ will be highly sensitive to how we handle
the bias issue, a point also raised in Ref.~\cite{Ichikawa:2006vm}.
We consider both a conservative and a more speculative
approach.

\subsection*{Conservative approach: LSS-lin}

In the conservative approach, we use power spectrum data only on
scales that are safely linear,
\begin{itemize}
\item 2dF-lin, $k_{\rm max} \sim 0.09 \ h \ {\rm Mpc}^{-1}$ (17 bands),
\item SDSS-DR2-lin, $k_{\rm max} \sim 0.06 \ h \ {\rm Mpc}^{-1}$ (11 bands), and
\item SDSS-LRG-lin from DR5, $k_{\rm max} \sim 0.07 \ h \ {\rm Mpc}^{-1} $ (11 bands).
\end{itemize}
The combined set of these data is denoted LSS-lin. We adopt the
constant bias assumption (\ref{eq:constantbias}) for each data set,
and marginalise over each of the three bias parameters $b^2$ with a
flat prior.

\subsection*{Speculative approach: LSS-Q}

In the speculative approach, we use data sets collectively
denoted as LSS-Q that include
\begin{itemize}
\item 2dF-Q, $k_{\rm max} \sim 0.15 \ h \ {\rm Mpc}^{-1}$ (32 bands),
\item SDSS-DR2-Q, $k_{\rm max} \sim 0.1 \ h \ {\rm Mpc}^{-1}$ (14 bands), and
\item SDSS-LRG-Q from DR5, $k_{\rm max} \sim 0.2 \ h \ {\rm Mpc}^{-1} $ (20
bands),
\end{itemize}
with $k_{\rm max}$ values chosen to conform with
the analyses of Refs.~\cite{Tegmark:2006az} and~\cite{Spergel:2006hy}. 
Here, we use the bias correction formula (\ref{eq:fittingformula})
and marginalise over each set of $b^2$ and $Q_{\rm nl}$  with flat
priors.%
\footnote{Some recent analyses use a Gaussian prior of $Q_{\rm nl}=4.6\pm1.5$
when fitting the 2dF data.  We point out that these numbers are
in fact {\it derived} from the 2dF data itself \cite{Cole:2005sx}.
We feel it is inconsistent to feed them back into a fit as a prior.}
Our motivation for caution in this case owes itself to the
fact that the formula (\ref{eq:fittingformula}) was originally
developed and calibrated for $\Lambda$CDM cosmologies; there is {\it
a priori} no guarantee that it would apply also to nonstandard
models.

We note that Seljak et~al.~\cite{Seljak:2006bg} and Mangano
et~al.~\cite{Mangano:2006ur} also used the bias correction formula
(\ref{eq:fittingformula}) on the SDSS-DR2 data.  However, they adopted
a Gaussian prior on $Q_{\rm nl}$ of $10 \pm 5$ that is predetermined
from numerical simulations. As we shall see, this choice tends to bias
their results towards large values of $N_{\rm eff}$. We believe
this is the main origin of the discrepant  $N_{\rm eff}$ values reported
by different groups.

\section{Statistical inference}                  \label{sec:inference}

\subsection{Bayesian inference}

We use standard Bayesian inference techniques, and explore the
model parameter space with Monte Carlo Markov Chains (MCMC) generated using
the publicly available {\sc CosmoMC}
package~\cite{Lewis:2002ah,cosmomc}.

Given a set
of data ${\bm x}$, a direct probabilistic interpretation for
the degree of belief in the parameters ${\bm \theta}$ of an assumed
underlying model is given by the posterior probability distribution
\begin{equation}
 P({\bm \theta}|{\bm x})
 \propto L({\bm x}|{\bm \theta}) \ \pi({\bm \theta}).
\end{equation}
Here, the likelihood function $L({\bm x}|{\bm \theta})$ quantifies the
agreement of the data with an assumed set of parameter values,
while the prior probability $\pi({\bm \theta})$
represents our belief in what the true parameter values should be
before any data is taken. This inherent subjectivity of Bayesian
inference is a point of much criticism.  A pragmatic approach
is to employ uniform priors and ``let the data decide''.  However,
this approach is not entirely free of subjectivity, particularly
when it comes to credible interval construction
and marginalisation (section~\ref{sec:marge}).

\subsection{Point estimates}

The posterior probability $P({\bm \theta}|{\bm x})$ serves as the
starting point for any further inference. A natural point of reference
is the posterior mode
\begin{equation}
 \hat{\bm \theta} =
 \arg \left[\max_{\bm \theta} P({\bm \theta}|{\bm x}) \right],
\end{equation}
representing the most probable parameter values given the data and
priors.  Note that we sometimes refer to the posterior mode as
the ``best-fit'', although strictly speaking the term refers to those
parameter values that maximise the likelihood and is equivalent to
$\hat{\bm \theta}$ only for
uniform priors. Another commonly used point estimate is the
posterior mean or ``expectation value''
\begin{equation}\label{eq:mean}
 \langle {\bm \theta} \rangle =
 \int d {\bm \theta} \ {\bm \theta} \ P({\bm \theta}|{\bm x}).
\end{equation}
For one-dimensional distributions, one may also define the median
$\theta_{\rm med}$, where 50\% of the posterior's volume lie on
either side.

\subsection{Credible intervals}
\label{sec:ci}

\begin{figure}
\hspace{25mm}
\includegraphics[width=9.cm]{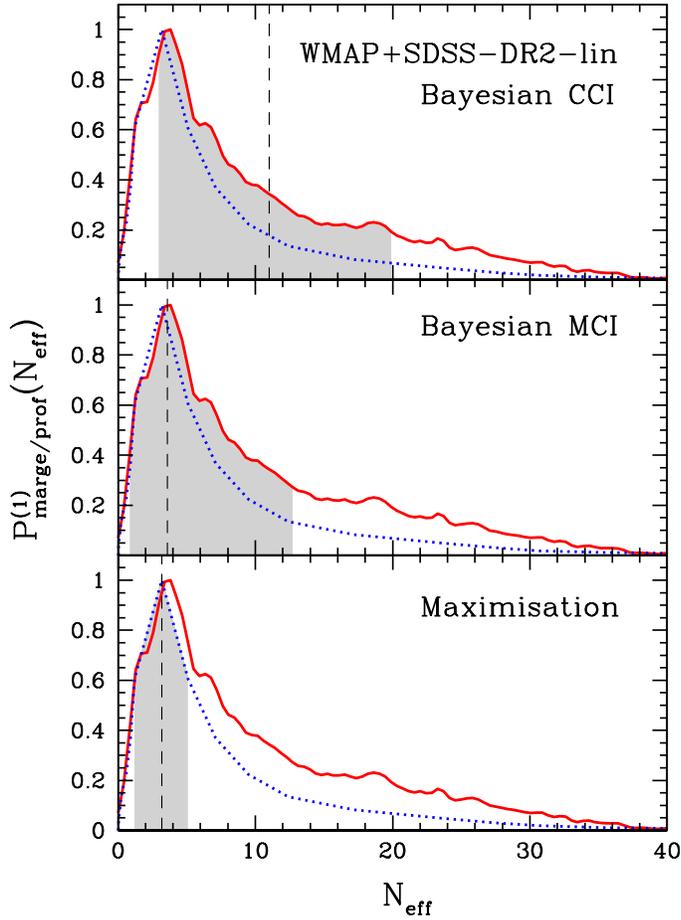}
\caption{The 1D marginal (red/solid) and profile (blue/dotted)
posteriors with respect to $N_{\rm eff}$ for our minimal model, the
data set WMAP+SDSS-DR2-lin and top hat prior $0.2 \leq  h \leq 2.0$.
The shaded regions are, from top to bottom, the Bayesian 68\%
central credible interval, the 68\% minimum credible interval, and
the $1\sigma$ interval derived from maximisation.  The dashed
vertical lines mark, from top to bottom, the posterior mean $\langle
N_{\rm eff} \rangle$, the 1D marginal posterior mode $\hat{N}_{\rm
eff}^{(1)}$, and the global best~fit~$\hat{N}_{\rm eff}$.
\label{fig:intervals}}
\end{figure}

In addition to point estimates one needs credible regions in parameter
space that express the degree of uncertainty in the inference. A
closed but not necessarily connected hypersurface $\partial A_\gamma$,
called a $100 \gamma$\% credible region, can be constructed such that
the hypervolume $A_\gamma$ contains a fraction $\gamma$ of the total
volume beneath $P({\bm \theta}|{\bm x})$,
\begin{equation}
\label{eq:hypersurface}
\int_{A_\gamma} d {\bm \theta} \ P({\bm \theta}|{\bm x}) = \gamma.
\end{equation}
This definition is not unique. In the 1D case, two popular choices are
\begin{itemize}
\item
{\it Central credible interval (CCI)}\quad The credible interval
$[\theta_{\rm lo},\theta_{\rm hi}]$ means that equal fractions
$(1-\gamma)/2$ of the posterior's volume lie in
$(-\infty,\theta_{\rm lo})$ and $(\theta_{\rm hi},\infty)$. The CCI
is always connected and contains the median $\theta_{\rm med}$.
\item
{\it Minimum credible interval (MCI)}\quad For a unimodal
distribution, $\theta_{\rm lo}$ and $\theta_{\rm hi}$ are chosen to
minimise $\theta_{\rm hi}\!-\theta_{\rm lo}$. This amounts to
placing $[\theta_{\rm lo},\theta_{\rm hi}]$ around the peak of the
posterior. In general the posterior may be multimodal, and the MCI is
constructed such that the posterior at any point inside is larger
than that at any point outside. The MCI need not be connected, but always
includes the mode $\hat\theta$.
\end{itemize}
These constructions coincide only under special circumstances, e.g.,
if the posterior probability is Gaussian with respect to $\theta$.
The top two panels of
figure~\ref{fig:intervals} show realistic examples of a CCI and an
MCI that are very different.

Which of these constructions should we adopt? Since our goal is to
find the most probable set of parameter values, we believe that the
MCI is more adequate because it singles out regions of parameter
space with the highest probability densities. In particular, the MCI
always includes the ``best-fit'' parameter (more precisely, the
mode). Finally, for multidimensional posteriors, only the MCI is
uniquely defined.

We discuss these matters in such detail because {\sc CosmoMC}'s
popular companion package {\sc GetDist}
outputs for 1D intervals a CCI, not an MCI, a property that does not
always seem to be recognised. Moreover, under the default settings,
{\sc GetDist} does not output the median $\theta_{\rm med}$,
the point estimate naturally associated with the~CCI, but rather the
expectation value $\langle \theta \rangle$.

\subsection{Marginalisation of the posterior}
\label{sec:marge}

For multi-parameter models typically encountered in cosmology,
the information carried by the multi-dimensional hypersurface
$\partial A_\gamma$ is often not useful in practice and must be
``compressed.'' It is common to map the posterior probability
$P({\bm \theta}|{\bm x})$ onto a lower-dimensional subspace by the
process of marginalisation,
\begin{equation}\label{eq:marginalisation}
 P^{(n)}_{\rm marge}({\bm \theta}^{(n)}) \propto
 \int d\theta_{n+1} \ldots d\theta_{N} \ P({\bm \theta}|{\bm x}),
\end{equation}
where ${\bm \theta}^{(n)}=(\theta_1,\dots,\theta_n)$ represents the
parameters in the $n$-dimensional subspace. Point estimates for ${\bm
\theta}^{(n)}$ and credible regions may then be constructed from the
marginal posterior probability in analogy to section~\ref{sec:ci}
above.

Marginalisation favours regions of parameter space that contain a
large volume of the probability density in the marginalised
directions. This ``volume effect'' can sometimes lead to counter-intuitive
results, such as suppression of the probability density for
the global best fit parameters $\hat{\bm
\theta}$ if they appear within sharp peaks or ridges that contain
little volume. Moreover, the concept of volume itself depends on the
choice of parameters. For example, a flat prior on a parameter or one
on its logarithm have completely different effects on the volume in
that parameter direction. Therefore, other methods of mapping the
multi-dimensional posterior
onto a lower-dimensional space can be
useful.

\subsection{Maximisation of the posterior}

A complementary approach to marginalisation
 is to project $P({\bm \theta}|{\bm
x})$ onto the $n$-dimensional subspace ${\bm \theta}^{(n)}$ by
maximising along the remaining directions,
\begin{equation}
P^{(n)}_{\rm prof}({\bm \theta}^{(n)}) \propto
\max_{\theta_{n+1},\ldots,\theta_{N}}  P({\bm \theta}|{\bm x}).
\end{equation}
The resulting $n$-dimensional profile posterior $P^{(n)}_{\rm
prof}({\bm \theta}^{(n)})$ has the advantage of preserving the true
peak of the original $N$-dimensional posterior probability and hence
the global best fit~$\hat{\bm \theta}$.
Figure~\ref{fig:intervals} shows a
realistic example of a 1D marginal and a 1D profile posterior in
juxtaposition.

In addition, we introduce an effective chi-square measure for the
goodness-of-fit relative to the global best fit,
\begin{equation}\label{eq:deltachi2}
 \Delta \chi^2_{\rm eff}({\bm \theta}^{(n)})
 \equiv - 2 \ln\left[
 \frac{P^{(n)}_{\rm prof}({\bm \theta}^{(n)})}
 {P(\hat{\bm \theta}|{\bm x})}\right].
\end{equation}
For $n=1$, we define loosely the ``$1 \sigma$'' and ``$2 \sigma$''
intervals as the 1D regions satisfying respectively $\Delta
\chi^2_{\rm eff} \leq 1$ and $\Delta \chi^2_{\rm eff} \leq 4$.  We
emphasise that these intervals have no formal probabilistic
interpretation. However, they do provide a raw assessment,
unplagued by volume effects, of how well a given parameter value
agrees with the data relative to the global best fit, and have
the virtue of being invariant
under reparameterisation of the model.
 Of course, if
$P^{(1)}_{\rm prof/marge}(\theta)$ is Gaussian, then the $1 \sigma$
and $2 \sigma$ intervals thus derived coincide with the 1D marginal
68\% and 95\% minimum and central credible
regions~\cite{Tegmark:2000db}.
Maximisation was used in some recent studies of cosmological
$N_{\rm eff}$ inference \cite{Crotty:2004gm,Hannestad:2003ye,Hannestad:2006mi,%
Cirelli:2006kt,Ichikawa:2006vm}.

For simplicity our maximisation intervals are extracted from the same MCMC 
chains used to construct the Bayesian credible intervals.  However, we caution that 
MCMC techniques are strictly speaking not designed for this purpose; 
there exist sophisticated optimisation methods such as simulated annealing 
that are much better suited to the task.

The bottom panel of figure~\ref{fig:intervals} shows a realistic
example of a one-dimensional $1 \sigma$ interval constructed
according to equation~(\ref{eq:deltachi2}). For a very non-Gaussian
situation such as depicted in this figure, the point estimates and
corresponding credible intervals derived by the methods discussed here
are very different.

\section{Constraints in the minimal model}
\label{sec:minimal}

\subsection{Numerical results}

To study the impact of different statistical methodologies and of
different combinations of data sets, we use the minimal model (i.e.,
vanilla+$N_{\rm eff}$) as a benchmark case. Each entry in
table~\ref{table:modela} gives a point estimate and the lower and
upper ends of the appropriate 68\% and 95\% credible intervals for
$N_{\rm eff}$. The first column indicates the combinations of
cosmological data sets. To illustrate the strong degeneracy between
$N_{\rm eff}$ and the Hubble parameter $h$ in some data sets and its
consequences, we have used two different top-hat priors: the loose
prior~1 ($0.2 \leq h \leq 2.0$) and the more constraining prior~2
($0.4 \leq h \leq 1.0$).

In the columns showing the Bayesian central credible interval, we
use the posterior mean $\langle N_{\rm eff}\rangle$ as a point
estimate, which is the default output of {\sc GetDist}. The
Bayesian minimum credible interval is derived from the 1D marginal posterior
probability distribution for $N_{\rm eff}$ and the corresponding point
estimate is the 1D marginal posterior mode $\hat N_{\rm eff}^{(1)}$. In the
case of maximisation, the point estimate is the global best fit
$\hat N_{\rm eff}$.   Here, the associated intervals are the
effective $1\sigma$ and $2\sigma$ regions defined by
equation~(\ref{eq:deltachi2}).

\begin{table}[t]
\caption{Point estimates and credible intervals (68\% and 95\%) for
$N_{\rm eff}$ in our minimal model ``vanilla+$N_{\rm eff}$''. The priors for the free
parameters are given in table~\ref{tab:priors}. Priors~1 and~2 differ
only for the Hubble parameter.  We consider also two large combinations
of data sets, \hbox{All-lin = WMAP+BAO+SNIa+LSS-lin} and \hbox{All-Q =
WMAP+BAO+SNIa+LSS-Q.}\label{table:modela}}
 \hbox to\hsize{\hfil
{\footnotesize
\begin{tabular}{@{}lcccccc}
\br
& \centre{2}{Bayesian CCI}  & \centre{2}{Bayesian MCI}  & \centre{2}{Maximisation} \\
\ms
& \centre{2}{$\langle N_{\rm eff} \rangle^{~68 \% \uparrow,~95 \% \uparrow}_{~68 \% \downarrow,~95 \% \downarrow}$}
& \centre{2}{$\hat{N}_{\rm eff}^{(1)}
\vphantom{\hat{N}_{\rm eff}}^{~68 \% \uparrow,~95 \% \uparrow}_{~68 \% \downarrow,~95 \% \downarrow}$}
& \centre{2}{$\hat{N}_{\rm eff}
\vphantom{\hat{N}_{\rm eff}}^{~1 \sigma \uparrow,~2 \sigma \uparrow}_{~1 \sigma \downarrow,~2 \sigma \downarrow}$} \\
\ns
& \crule{2} & \crule{2} & \crule{2}  \\
Data  & Prior 1 & Prior 2 & Prior 1 & Prior 2 & Prior 1 & Prior 2 \\
\br
\ms
WMAP & $22^{~37,~46}_{~7.3,~2.6}$
&  $5.8^{~8.8,~11}_{~3.0,~1.5}$
& $6.8^{~32,~45}_{~2.8,~1.5}$
& $4.2^{~7.9,~11}_{~2.2,~1.2}$
& $3.9^{~6.1,~27}_{~1.5,~0.6}$
& $3.9^{~6.1,~12}_{~1.5,~0.6}$
\\
\ms
\mr
\ms
+SDSS-DR2-Q  & $14^{~26,~37}_{~3.6,~1.2}$
& $4.8^{~7.7,~10}_{~2.1,~1.0}$
& $3.6^{~18,~34}_{~0.6,~0.0}$
& $3.7^{~6.4,~9.7}_{~1.1,~0.7}$
& $2.6^{~5.6,~14}_{~1.0,~0.3}$
& $2.3^{~5.6,~11}_{~1.0,~0.6}$
\\
\bs
+SDSS-DR2-lin  &  $11^{~20,~32}_{~3.0,~1.2}$
& $4.9^{~8.0,~10}_{~2.0,~0.7}$
& $3.6^{~13,~28}_{~0.7,~0.3}$
& $4.3^{~6.5,~9.9}_{~0.9,~0.5}$
& $3.2^{~5.1,~12}_{~1.2,~0.2}$
& $3.2^{~5.1,~11}_{~1.3,~0.2}$
\\
\bs
+2dF-Q  &  $3.2^{~5.2,~8.4}_{~1.1,~0.3}$
& $2.6^{~4.3,~5.7}_{~1.1,~0.4}$
& $1.6^{~4.2,~7.5}_{~0.4,~0.0}$
& $1.4^{~3.9,~5.5}_{~0.7,~0.0}$
& $1.5^{~2.4,~5.3}_{~0.6,~-}$
& $1.5^{~2.4,~5.0}_{~0.6,~-}$
\\
\bs
+2dF-lin &  $4.6^{~7.1,~10}_{~2.2,~1.1}$
& $4.4^{~6.8,~9.6}_{~2.1,~1.1}$
& $2.9^{~5.8,~9.5}_{~1.3,~0.6}$
&  $3.2^{~5.7,~9.4}_{~1.4,~0.7}$
& $2.6^{~4.5,~7.9}_{~1.2,~0.6}$
& $2.6^{~4.5,~7.9}_{~1.2,~0.6}$
\\
\bs
+SDSS-LRG-Q & $3.5^{~5.1,~7.4}_{~2.0,~1.1}$
& $3.5^{~5.1,~7.4}_{~2.0,~1.1}$
& $2.6^{~4.5,~6.9}_{~1.5,~0.8}$
& $2.5^{~4.5,~6.9}_{~1.5,~0.8}$
&  $2.7^{~4.1,~6.3}_{~1.5,~0.8}$
&  $2.7^{~4.1,~6.3}_{~1.5,~0.8}$
\\
\bs
+SDSS-LRG-lin & $4.0^{~5.8,~9.4}_{~2.1,~1.2}$
& $3.5^{~5.0,~6.6}_{~2.1,~1.3}$
& $2.6^{~4.9,~8.4}_{~1.5,~0.7}$
& $2.8^{~4.6,~6.3}_{~1.8,~1.1}$
& $2.7^{~4.3,~6.2}_{~1.8,~0.8}$
& $2.7^{~4.0,~6.2}_{~1.8,~1.2}$
\\
\bs
+BAO & $3.5^{~5.0,~6.8}_{~2.1,~1.1}$
& $3.5^{~5.0,~6.8}_{~2.1,~1.1}$
& $2.8^{~4.7,~6.4}_{~1.8,~0.8}$
& $2.8^{~4.7,~6.4}_{~1.8,~0.8}$
& $2.1^{~4.7,~6.6}_{~1.4,~0.9}$
&  $2.1^{~4.7,~6.6}_{~1.4,~0.9}$
\\
\bs
+SNIa & $20^{~34,~44}_{~6.4,~2.3}$
& $5.9^{~9.1,~11}_{~2.8,~0.9}$
& $4.3^{~28,~42}_{~2.8,~0.4}$
& $4.1^{~8.7,~11}_{~2.4,~0.9}$
& $3.6^{~6.3,~24}_{~1.4,~0.3}$
& $3.6^{~6.3,~12}_{~1.6,~0.3}$
\\
\bs
+HST & $3.9^{~5.7,~8.3}_{~2.1,~1.2}$
&  $4.0^{~5.7,~7.5}_{~2.4,~1.4}$
&  $3.3^{~5.1,~7.7}_{~1.6,~0.8}$
& $3.6^{~5.3,~7.0}_{~2.1,~1.0}$
& $2.9^{~4.6,~7.6}_{~1.6,~0.4}$
& $2.9^{~4.5,~6.4}_{~1.6,~0.9}$
\\
\bs
+Ly$\alpha$ & $7.6^{~10,~13}_{~5.2,~3.6}$
& $6.9^{~9.0,~11}_{~4.9,~3.5}$
& $6.8^{~9.3,~12}_{~4.6,~3.3}$
& $6.4^{~8.8,~11}_{~4.6,~3.2}$
& $6.6^{~8.0,~12}_{~4.9,~3.3}$
& $6.6^{~7.7,~10}_{~5.3,~3.3}$
\\
\mr
\ms
All-lin
& ---
&  $2.9^{~4.0,~5.3}_{~1.8,~1.1}$
& ---
& $2.6^{~3.7,~5.1}_{~1.5,~0.9}$
& ---
& $2.7^{~3.3,~5.0}_{~1.5,~0.8}$ \\
\bs
All-lin+HST
& ---
&  $2.8^{~3.7,~4.9}_{~1.9,~1.3}$
& ---
& $2.6^{~3.6,~4.8}_{~1.8,~1.1}$
& ---
& $2.7^{~3.2,~4.5}_{~2.0,~1.1}$\\
\bs
All-Q
& ---
&  $2.3^{~3.2,~4.4}_{~1.4,~0.7}$
& ---
& $2.0^{~3.1,~4.1}_{~1.2,~0.5}$
& ---
& $2.0^{~2.4,~4.0}_{~1.3,~0.6}$ \\
\bs
All-Q+HST
& ---
&  $2.5^{~3.5,~4.3}_{~1.6,~1.0}$
& ---
&  $2.4^{~3.3,~4.3}_{~1.6,~0.9}$
& ---
& $2.2^{~2.7,~3.8}_{~1.6,~0.9}$ \\
\bs
All-Q+Ly$\alpha$
& ---
&  $4.4^{~5.5,~6.9}_{~3.3,~2.4}$
& ---
& $4.4^{~5.4,~6.6}_{~3.2,~2.3}$
& ---
& $4.2^{~4.7,~6.4}_{~3.4,~2.4}$ \\
\bs
All-Q+Ly$\alpha$+HST
& ---
& $3.9^{~4.8,~5.9}_{~3.0,~2.3}$
& ---
& $3.8^{~4.7,~5.8}_{~2.9,~2.2}$
& ---
& $4.0^{~4.3,~5.6}_{~3.1,~2.3}$ \\
\br
\end{tabular}}}
\end{table}

\subsection{Interpretation of statistics}

To compare estimates from different inference schemes, consider
first the top half of table~\ref{table:modela}.
The posterior mean and the CCI, i.e., the default output
of {\sc GetDist},
show a preference for large $N_{\rm eff}$ for almost all
combinations of probes. The combinations WMAP, WMAP+SDSS-DR2, and
WMAP+SNIa, in particular, appear to disfavour the standard value $N_{\rm
eff}=3.046$ at more than 68\% (prior~1). However, any evidence for
$N_{\rm eff}>3.046$ disappears as soon as we impose the tighter
prior~2 on $h$. This trend stems from the $N_{\rm
eff}$-$h$-degeneracy which leads to a long tail of high $N_{\rm eff}$
in the 1D marginal posterior (figure~\ref{fig:intervals}).  The tail
in turn pushes the posterior
mean and the
CCI to larger $N_{\rm eff}$ values. Imposing a tighter prior on $h$
suppresses the tail and reduces this effect.

In contrast, the 1D marginal posterior mode $\hat{N}_{\rm
eff}^{(1)}$ and the global best fit  $\hat{N}_{\rm eff}$ pick out the
parameters with the highest probability densities, and turn out to
be insensitive to the choice of $h$ prior. The tail region still has
a strong impact on the upper MCI limits, but the lower limits are
relatively unaffected. The $N_{\rm eff}$ constraints from WMAP in
table \ref{table:modela} provide an excellent illustration of this
point.

The $1 \sigma$ and $2 \sigma$ intervals from maximisation depend
even less on the $h$ prior, since this construction makes no
reference to the volume of the posterior and is therefore
insensitive to tail regions once the 1D profile posterior drops
below $e^{-2}$ relative to the peak. As argued earlier
(section~\ref{sec:ci}), in Bayesian inference only the MCI provides
a meaningful answer to the question, what are the most probable
values of $N_{\rm eff}$ implied by the data. Our explicit examples
show that inference based on the CCI, the default output of {\sc
GetDist}, can lead to incorrect conclusions.

\subsection{Scale-dependent bias}

Turning to the issue of bias in the galaxy power spectrum, we see in
table~\ref{table:modela} that the two different measures introduced
in section~\ref{sec:bias} to bypass or account for the scale
dependence, namely, using only linear data at $k < 0.1 \ h \ {\rm
Mpc}^{-1}$, or adopting the bias correction  formula
(\ref{eq:fittingformula}), generally produce consistent results.
The agreement between WMAP+SDSS-DR2-lin and
WMAP+SDSS-DR2-Q, and between WMAP+SDSS-LRG-lin and WMAP+SDSS-LRG-Q
are excellent, suggesting that the effects of scale-dependent biasing
have been successfully ameliorated. The WMAP+2dF-lin and WMAP+2dF-Q
results do show a slight discrepancy at roughly the 68\% level. This
can most likely be put down to statistical fluctuations, but recall
that the bias correction formula (\ref{eq:fittingformula}) has not
been tested for nonstandard cosmologies and its application here is, strictly speaking,
experimental.

The analyses of Seljak et~al.~\cite{Seljak:2006bg} and Mangano
et~al.~\cite{Mangano:2006ur} found a very high $N_{\rm
eff}=7.8^{~8.9,~10}_{~7.1,~4.6}$ for WMAP+SDSS-DR2+SNIa,
which can only be accommodated within our corresponding
MCI estimates,
$N_{\rm eff}=3.7^{~6.4,~9.7}_{~1.1,~0.7}$ for WMAP+SDSS-DR2-Q and $N_{\rm
eff}=4.3^{~6.5,~9.9}_{~0.9,~0.5}$ for WMAP+SDSS-DR2-lin, at more than
the 68\% level. Both groups
used the bias correction formula (\ref{eq:fittingformula}), but
adopted the Gaussian prior $Q_{\rm nl} = 10 \pm 5$, a range supposedly
determined from numerical simulations, although no source is
cited. As a test, we have performed a fit of WMAP+SDSS-DR2-Q+SNIa
using the same Gaussian prior on $Q_{\rm nl}$.
We find $N_{\rm eff}=6.2^{~10,~12}_{~4.1,~1.9}$
(MCI) and $N_{\rm eff}=7.0^{~9.9,~12}_{~4.1,~2.2}$ (CCI),
which include the high $N_{\rm eff}$ values of Refs.~\cite{Seljak:2006bg,Mangano:2006ur}
in the 68\% region.
Excluding SNIa from the fit yields essentially the same constraints.

These test results clearly
indicate that the choice of $Q_{\rm nl}$ prior plays an important role in the
inference of $N_{\rm eff}$.  In this case, the choice of $Q_{\rm nl}=10\pm5$ tends
to push the preferred $N_{\rm eff}$ to higher values.
We are not able to reproduce the very tight error bars for $N_{\rm eff}$
reported in Refs.~\cite{Seljak:2006bg,Mangano:2006ur}, which
may be due to different priors assumed for the marginalised parameters,
or because of a slightly larger $k_{\rm max} \sim 0.15 \ h \ {\rm Mpc}^{-1}$
adopted in these analyses.
However, we also observe a peculiar feature in their credible intervals:
the 68\% interval is some three times smaller than the 95\% interval.  This
suggests some highly non-Gaussian behaviour in their marginal posterior
for $N_{\rm eff}$, because in a Gaussian distribution, the ratio of the intervals
is $1:2$.

\begin{figure}
\hspace{25mm}
\includegraphics[width=11.5cm]{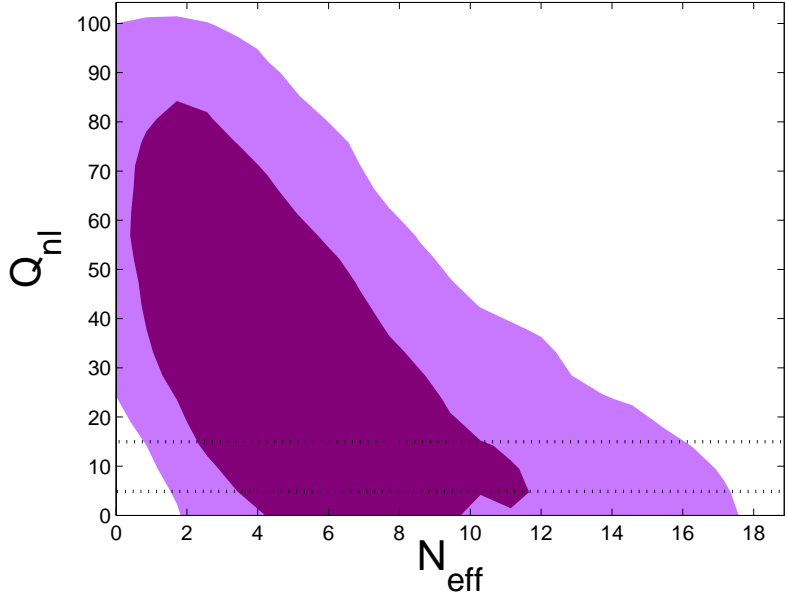}
\caption{The 2D marginal 68\% and 95\% allowed regions in the minimal
model for $N_{\rm eff}$ and $Q_{\rm nl}$, using the data set
WMAP+SDSS-DR2-Q+SNIa and prior 2.  The horizontal dotted lines indicate the
$1 \sigma$ range of the Gaussian
prior $Q_{\rm nl}=10\pm5$.  \label{fig:QN}}
\end{figure}

The dependence on the $Q_{\rm nl}$ prior traces its origin
to a degeneracy between $N_{\rm eff}$ and $Q_{\rm nl}$.
Figure~\ref{fig:QN} shows the 2D marginal 68\% and 95\% allowed regions
in $N_{\rm eff}$-$Q_{\rm nl}$-space for the data set WMAP+SDSS-DR2-Q+SNIa.
Evidently, imposing the restrictive
prior $Q_{\rm nl}=10\pm5$ cuts off much of the parameter space that favours
low values of $N_{\rm eff}$.
To our knowledge no simulation of mock galaxy catalogues involving a
nonstandard $N_{\rm eff}$ value has ever been reported in the literature.
Without the backing of simulations (or other independent input)
there is no justification to impose
a restrictive prior
on $Q_{\rm nl}$ when performing a fit with $N_{\rm eff}$ as a free parameter.
The best strategy in such circumstances is to use a broad and uniform prior on
$Q_{\rm nl}$, as adopted in our analysis and
also advocated in Ref.~\cite{Tegmark:2006az}.

To summarise, we find that imposing a $Q_{\rm nl}=10\pm5$ prior for
the WMAP+SDSS-DR2-Q+SNIa fit biases the preferred $N_{\rm eff}$ to
higher values.  This may account for the difference between our
result and those reported in
Refs.~\cite{Seljak:2006bg,Mangano:2006ur}.%
\footnote{For
completeness, we quote here the constraints on $Q_{\rm nl}$ derived
from WMAP+SDSS-DR2 using 19 data bands (i.e., $k_{\rm max} \sim 0.2 \ h \ 
{\rm Mpc}^{-1}$) 
in the vanilla
model: $Q_{\rm nl}=15^{+5}_{-4}$ (68\% C.L.). 
Here, five additional  
data points at large $k$ values allow one to place much tighter constraints on $Q_{\rm nl}$ than is
possible with only 14 data bands used in, e.g., figure~\ref{fig:QN}.
This result should be compared
with $Q_{\rm nl}=30^{+4.4}_{-4.1}$ for WMAP+SDSS-LRG (20~bands)~\cite{Tegmark:2006az} 
and $Q_{\rm nl}=4.6\pm1.5$ for WMAP+2dF (36~bands)
\cite{Cole:2005sx} for the same model.  }

\subsection{Combining all data sets}

Having identified and corrected the problematic issues, we now turn
to our own $N_{\rm eff}$ estimates. An inspection of
table~\ref{table:modela} reveals that, except for those sets
including Ly$\alpha$, none of the combinations of probes shows any
significant evidence for $N_{\rm eff} \neq 3.046$, a value that
always sits comfortably within the 68\% MCI. The combination of all
linear data together with HST (All-lin+HST) gives $N_{\rm
eff}=2.6^{~3.6,~4.8}_{~1.8,~1.1}$. Discarding HST leaves the best
fit unchanged, but slightly loosens the credible intervals.

Including nonlinear data in the galaxy power spectrum tends to
reduce the numbers a little to $N_{\rm
eff}=2.0^{~3.1,~4.1}_{~1.2,~0.5}$ (All-Q), essentially because 2dF-Q
prefers a low $N_{\rm eff}$. Adding HST shifts it up again to
$N_{\rm eff}=2.4^{~3.3,~4.3}_{~1.6,~0.9}$. We repeat that the bias
correction formula~(\ref{eq:fittingformula}) may not be applicable
in nonstandard cosmologies so that numbers from the Q sets must be
interpreted with caution.

Another interesting feature is that, with the exception of 
WMAP+2dF-Q, all combinations of data sets prefer a nonzero $N_{\rm eff}$
at the 95\% level or better.  This is in contrast to the results
of Ref.~\cite{Ichikawa:2006vm}, which finds no lower 95\% limit 
from the WMAP data alone.  We have not investigated where the 
differences come from.  As mentioned before, the WMAP+2dF-Q data set 
tends to prefer lower values of $N_{\rm eff}$ and as such produces 
no lower 95\% limit on $N_{\rm eff}$.

The Ly$\alpha$ data appear to be the only data set that prefers a much
larger value of $N_{\rm eff}$, with WMAP+Ly$\alpha$ disfavouring
$N_{\rm eff}=3.046$ at 95\%. When combined with other data sets, however,
the evidence against $N_{\rm eff} = 3.046$ is weakened to the  68\%
level, $N_{\rm eff} = 3.8^{~4.7,~5.8}_{~2.9,~2.2}$ for
All-Q+Ly$\alpha$+HST, because 2dF-Q's preference for small $N_{\rm
eff}$ values tends to pull in the opposite direction.

\begin{figure}
\hspace{25mm}
\includegraphics[width=12cm]{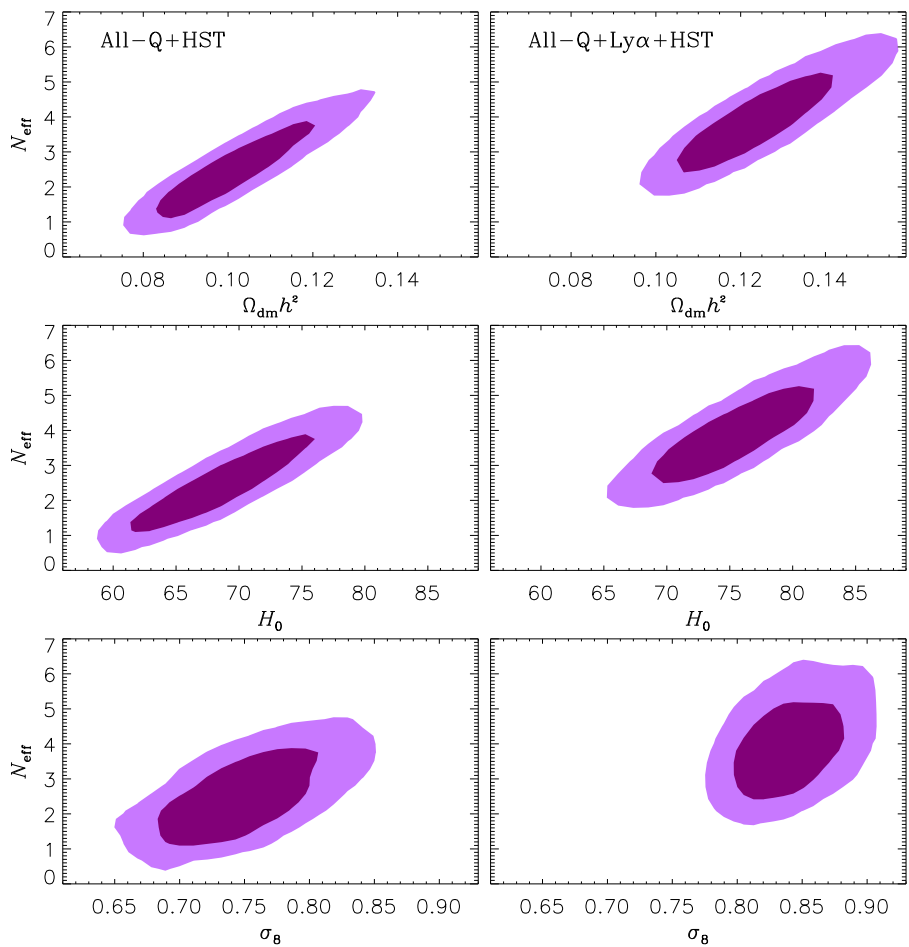}
\caption{The 2D marginal 68\% and 95\% allowed regions in the minimal
model for the indicated pairs of parameters. Plots in the left column use
the All-Q+HST data set, while those in the right column include also
Ly$\alpha$  (All-Q+Ly$\alpha$+HST). \label{fig:degeneracies}}
\end{figure}

The origin of Ly$\alpha$'s preference for large values of $N_{\rm eff}$ can be
gleaned from figure~\ref{fig:degeneracies}. The Ly$\alpha$ data
prefer a much higher amplitude of density fluctuations at small
scales, quantified by $\sigma_8$, than other data sets. This is
particularly evident in the bottom panels of
figure~\ref{fig:degeneracies}. The higher $\sigma_8$ value required by
Ly$\alpha$
forces $N_{\rm eff}$ upwards and cuts away the allowed region for low
$N_{\rm eff}$ values. As can be seen in the same figure,
with the inclusion of  Ly$\alpha$, the upper bound on
$N_{\rm eff}$ comes mainly from the HST
prior on $H_0$. Since $N_{\rm eff}$ and $H_0$ both control the epoch of
matter--radiation equality and are thus strongly degenerate,
a large $N_{\rm eff}$ can only be accommodated by a high
value of $H_0$. However, such high values are strongly disfavoured
by the HST data.

The overall shift in the allowed range for $N_{\rm eff}$ between
WMAP+Ly$\alpha$ and All-Q+Ly$\alpha$ also points to the fact that
the SDSS-Ly$\alpha$ data is not completely compatible with other
data sets (see, e.g., Refs.~\cite{Seljak:2006bg,Goobar:2006xz}).

\subsection{Towards Gaussianity}

A striking feature in table~\ref{table:modela} is that when all data
sets are combined, the three different statistical methods give almost
identical results. The reason is that the combination of CMB, LSS,
and SNIa data effectively breaks all parameter degeneracies and
yields a posterior distribution that is very close to Gaussian, a limit
in which all three methods must give the same result. The lower half of
table~\ref{table:modela} nicely confirms this expectation.

\section{Extended models}                         \label{sec:extended}

We now consider constraints on $N_{\rm eff}$ in the context of
extended models that allow also for nonvanishing neutrino masses. As in
the case of the minimal model, we calculate the bounds within a
conservative approach using only linear data (All-lin), as well as a
more speculative one that utilises the stronger, but more
model-dependent All-Q data set. Since, as we saw in
section~\ref{sec:minimal}, $N_{\rm eff}$ exhibits a strong
degeneracy with the Hubble parameter $H_0$ in some data sets, we
consider both options of including and excluding the HST data in our
analysis.  We do not use the Ly$\alpha$ data for the extended
models. Table~\ref{table:extended} shows our constraints on $N_{\rm
eff}$ for four choices of extended models: vanilla+$N_{\rm eff}$
extended with $f_\nu$, $f_\nu$+$\alpha_s$+$w$, $^3f_\nu$, and
$^3f_\nu$+$\alpha_s$+$w$.

\begin{table}[t]
\caption{ Point estimates and credible intervals (68\% and 95\%) for
$N_{\rm eff}$ in four extended model spaces. In the top segment, the
minimal vanilla+$N_{\rm eff}$ model is extended with $f_\nu$ and
$f_\nu$+$\alpha_s$+$w$ ($N_m=N_{\rm eff}$), while in the middle segment
 the extensions are $^3f_\nu$ and $^3f_\nu$+$\alpha_s$+$w$
($N_m=3.046$) as defined in section~\ref{sec:models}. 
The bottom segment contains results for the minimal model 
copied from table~\ref{table:modela}.
The priors for
the free parameters are given in table~\ref{tab:priors}. The columns
headed ``prior 2'' use a top hat prior \hbox{$0.4<h<1.0$}, while
those with ``+HST'' use in addition the HST result. The data sets
used are \hbox{All-lin = WMAP+BAO+SNIa+LSS-lin} and \hbox{All-Q =
WMAP+BAO+SNIa+LSS-Q.}\label{table:extended}}
 \hbox to\hsize{\hfil{\footnotesize
 \begin{tabular}{@{}llcccccc}
\br
 & & \centre{2}{Bayesian CCI}  & \centre{2}{Bayesian MCI}
& \centre{2}{Maximisation} \\
\ms
&& \centre{2}{$\langle N_{\rm eff} \rangle^{~68 \% \uparrow,~95 \% \uparrow}_{~68 \% \downarrow,~95 \% \downarrow}$}
& \centre{2}{$\hat{N}_{\rm eff}^{(1)}
\vphantom{\hat{N}_{\rm eff}}^{~68 \% \uparrow,~95 \% \uparrow}_{~68 \% \downarrow,~95 \% \downarrow}$}
& \centre{2}{$\hat{N}_{\rm eff}
\vphantom{\hat{N}_{\rm eff}}^{~1 \sigma \uparrow,~2 \sigma \uparrow}_{~1 \sigma \downarrow,~2 \sigma \downarrow}$} \\
\ns
&& \crule{2} & \crule{2} & \crule{2}  \\
Model & Data & Prior 2 & +HST & Prior 2 & +HST & Prior 2 & +HST \\
\br \ms +$f_\nu$ & All-lin
& $4.0^{~5.6,~8.2}_{~2.5,~1.5}$
& $3.7^{~4.9,~6.3}_{~2.6,~1.8}$
& $3.2^{~5.0,~7.8}_{~2.0,~1.1}$
& $3.6^{~4.7,~6.1}_{~2.4,~1.6}$
&$3.0^{~4.6,~6.2}_{~2.0,~1.1}$
& $3.7^{~4.1,~5.7}_{~2.5,~1.8}$
\\
\bs +$f_\nu$ & All-Q
& $3.6^{~5.0,~7.0}_{~2.2,~1.1}$
& $3.5^{~4.5,~5.8}_{~2.4,~1.7}$
& $2.9^{~4.7,~6.6}_{~1.9,~0.8}$
& $3.2^{~4.3,~5.6}_{~2.2,~1.5}$
& $3.2^{~3.8,~5.5}_{~2.1,~1.3}$
& $3.0^{~3.8,~5.3}_{~2.3,~1.6}$
\\
\bs +$f_\nu$+$\alpha_s$+$w$ & All-lin
& $3.7^{~5.3,~8.1}_{~2.0,~1.0}$
& $3.7^{~5.1,~6.6}_{~2.3,~1.4}$
& $3.1^{~4.9,~7.6}_{~1.6,~0.4}$
& $2.6^{~4.7,~6.4}_{~2.0,~1.2}$
&$2.5^{~3.2,~5.5}_{~1.5,~0.8}$
& $3.0^{~3.6,~5.5}_{~2.3,~1.1}$
\\
\bs +$f_\nu$+$\alpha_s$+$w$ & All-Q
& $3.3^{~4.9,~7.8}_{~1.8,~0.9}$
& $3.3^{~4.6,~6.3}_{~1.9,~1.0}$
& $2.3^{~4.2,~6.8}_{~1.3,~0.5}$
&$3.0^{~4.3,~6.1}_{~1.7,~0.8}$
& $2.6^{~3.0,~5.1}_{~1.5,~0.5}$
& $2.9^{~4.2,~5.1}_{~1.7,~1.0}$
\\
\mr
\ms +$^3f_\nu$ & All-lin
& $4.9^{~5.3,~8.0}_{~3.0,~3.0}$
& $4.4^{~4.8,~6.7}_{~3.0,~3.0}$
& $3.2^{~5.3,~8.0}_{~3.0,~3.0}$
& $3.2^{~4.8,~6.7}_{~3.0,~3.0}$
& $3.0^{~3.8,~5.7}_{~3.0,~3.0}$
& $3.0^{~3.9,~5.7}_{~3.0,~3.0}$
\\
\bs +$^3f_\nu$ & All-Q
& $4.4^{~4.6,~7.1}_{~3.0,~3.0}$
& $4.2^{~4.5,~6.1}_{~3.0,~3.0}$
& $3.0^{~4.6,~7.1}_{~3.0,~3.0}$
& $3.2^{~4.5,~6.1}_{~3.0,~3.0}$
& $3.0^{~3.9,~5.2}_{~3.0,~3.0}$
& $3.0^{~3.7,~5.0}_{~3.0,~3.0}$
\\
\bs +$^3f_\nu$+$\alpha_s$+$w$ & All-lin
& $5.1^{~5.3,~9.4}_{~3.0,~3.0}$
& $4.4^{~4.7,~6.7}_{~3.0,~3.0}$
& $3.0^{~5.3,~9.4}_{~3.0,~3.0}$
& $3.5^{~4.7,~6.7}_{~3.0,~3.0}$
& $3.0^{~3.9,~6.4}_{~3.0,~3.0}$
& $3.2^{~4.0,~6.0}_{~3.0,~3.0}$
\\
\bs +$^3f_\nu$+$\alpha_s$+$w$ & All-Q
& $4.4^{~4.7,~7.3}_{~3.0,~3.0}$
& $4.1^{~4.3,~5.8}_{~3.0,~3.0}$
& $3.0^{~4.7,~7.3}_{~3.0,~3.0}$
& $3.2^{~4.3,~5.8}_{~3.0,~3.0}$
& $3.0^{~3.7,~5.0}_{~3.0,~3.0}$
& $3.0^{~3.8,~5.0}_{~3.0,~3.0}$
\\
\mr
\ms Minimal & All-lin
&  $2.9^{~4.0,~5.3}_{~1.8,~1.1}$
&  $2.8^{~3.7,~4.9}_{~1.9,~1.3}$
& $2.6^{~3.7,~5.1}_{~1.5,~0.9}$
& $2.6^{~3.6,~4.8}_{~1.8,~1.1}$
& $2.7^{~3.3,~5.0}_{~1.5,~0.8}$ 
& $2.7^{~3.2,~4.5}_{~2.0,~1.1}$
\\
\bs Minimal & All-Q
&  $2.3^{~3.2,~4.4}_{~1.4,~0.7}$
& $2.5^{~3.5,~4.3}_{~1.6,~1.0}$
& $2.0^{~3.1,~4.1}_{~1.2,~0.5}$
& $2.4^{~3.3,~4.3}_{~1.6,~0.9}$
& $2.0^{~2.4,~4.0}_{~1.3,~0.6}$ 
& $2.2^{~2.7,~3.8}_{~1.6,~0.9}$
\\
\br
\end{tabular}}}
\end{table}

\subsection{$N_m=N_{\rm eff}$}
\label{sec:nm=neff}

Consider first the top half of table~\ref{table:extended}.  The
two extended models have, respectively, vanilla+$N_{\rm eff}$+$f_\nu$
and vanilla+$N_{\rm eff}$+$f_\nu$+$\alpha_s$+$w$ as free parameters.
Also in place is the condition \mbox{$N_m=N_{\rm eff}$}, meaning that
all $N_{\rm eff}$ neutrinos have equal masses $m_\nu$.
In both cases, it is evident that some new degeneracies have arisen
with the introduction of additional free parameters; the marginal
posteriors for $N_{\rm eff}$ are not perfect Gaussians
for the All-lin and All-Q data sets, as indicated by the fact that
their associated credible intervals from different constructions do
not exactly overlap.  However, none of the All-lin and All-Q
results show any significant deviation from the standard $N_{\rm eff}=3.046$,
and adding the HST data essentially serves to tighten the bounds.

It is interesting to note that, in the case of the
smaller vanilla+$N_{\rm eff}$+$f_\nu$ model, adding the HST data
brings the marginal posterior for $N_{\rm eff}$ much closer to the
Gaussian limit, so that the three different credible interval construction
methods give almost identical results.   Our best estimate
is $N_{\rm eff}=3.2^{~4.3,~5.6}_{~2.2,~1.5}$ (All-Q+HST),
values that are somewhat larger than
those found in the minimal vanilla+$N_{\rm eff}$ model for the same
data set, $N_{\rm eff}=2.4^{~3.3,~4.3}_{~1.6,~0.9}$,
because of a degeneracy between $N_{\rm eff}$ and $f_\nu$.

For the even larger vanilla+$N_{\rm eff}$+$f_\nu$+$\alpha_s$+$w$ model,
an additional degeneracy between $N_{\rm eff}$ and $w$ comes into play
so that the posterior for $N_{\rm eff}$ becomes more non-Gaussian.
For All-Q+HST, for example, even though the MCI and the CCI have
more or less converged (thus indicating a symmetric marginal posterior),
the limits from maximisation are still very different.
As our formal bound we use the MCI estimate for All-Q+HST,
$N_{\rm eff}=3.0^{~4.3,~6.1}_{~1.7,~0.8}$, but also note that all three
methods give credible intervals that are compatible with $N_{\rm eff}=3.046$
at better than 68\%. Thus, as was the case for the minimal model,
there is no evidence for any nonstandard value of $N_{\rm eff}$.

\begin{figure}
\hspace{25mm}
\includegraphics[width=11.5cm]{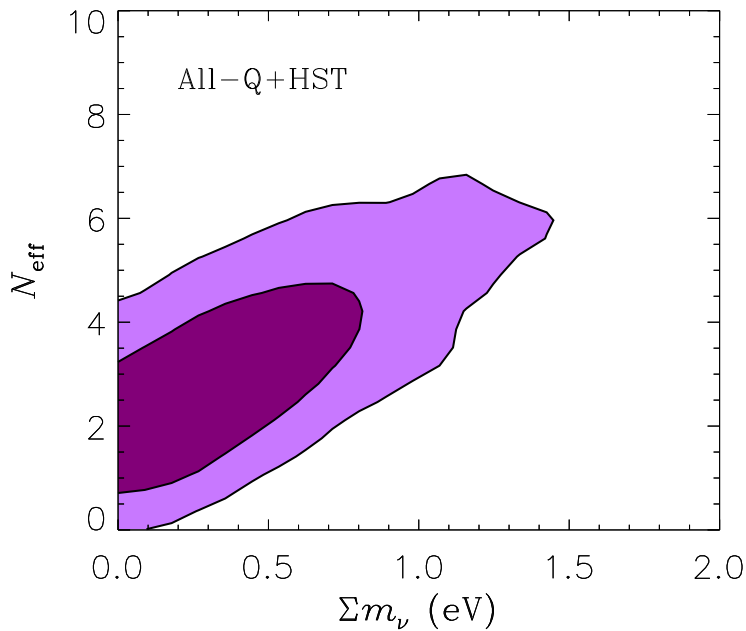}
\caption{The 2D marginal 68\% and 95\% allowed regions in $\sum m_\nu$ and
$N_{\rm eff}$ in the extended model vanilla+$N_{\rm eff}$+$f_\nu$+$\alpha_s$+$w$, using
the data set All-Q+HST.  The corresponding contours for the model 
vanilla+$N_{\rm eff}$+$^3f_\nu$+$\alpha_s$+$w$ are similar, but with a cut-off at $N_{\rm eff}=3$.
\label{fig:mnuNnu}}
\end{figure}

Figure~\ref{fig:mnuNnu} shows the 2D marginal contours in the  
$\sum m_\nu$-$N_{\rm eff}$ plane for the extended model 
vanilla+$N_{\rm eff}$+$f_\nu$+$\alpha_s$+$w$ and the data set All-Q+HST.
Some degeneracy persists between $\sum m_\nu$ and $N_{\rm eff}$, 
in contrast to earlier results from some of us \cite{Hannestad:2006mi}.  The 
difference can be traced to a generally more conservative approach taken in 
the present work, particularly with regard to scale-dependent biasing, as well as 
a different statistical methodology (Bayesian marginalisation vs maximisation).

\subsection{$N_m=3$}

The bottom half of table~\ref{table:extended} shows constraints on
$N_{\rm eff}$ for essentially
the same two classes of models, vanilla+$N_{\rm eff}$+$^3f_\nu$
and vanilla+$N_{\rm eff}$+$^3f_\nu$+$\alpha_s$+$w$,
except we now impose the condition $N_m=3$, representing
models with three massive neutrinos and $N_{\rm eff}-N_m$ massless
species.
This model is different from that presented above in section~\ref{sec:nm=neff}
because there is now a hard lower limit of $N_{\rm eff}=3$.

The presence of a hard limit can in principle lead to some
very disparate credible intervals from the three different
construction methods.  In the present case, however, the
1D marginal and profile posteriors for $N_{\rm eff}$ both
peak at or very near the limit.
It is therefore more useful to
report, instead of a CCI, an upper 100$\gamma$\% limit
constructed by requiring that a fraction $\gamma$ of the marginal
posterior's volume lies to the left of the limit.  This construction
is also a default setting of {\sc GetDist} for parameter estimation in
the presence of hard limits.  For simplicity, however, we shall
continue to label an interval thus constructed as a CCI.
The definitions of an MCI and a maximisation interval are the same as before.

The fact that the marginal posterior for $N_{\rm eff}$ peaks at or
very near the hard limit also means that, although the posterior
mean and mode still differ, the CCI and the MCI
will coincide, as is clearly shown in the bottom half of
table~\ref{table:extended}.  All estimates indicate that
$N_{\rm eff}=3.046$ sits safely within the 68\% region.
Our best estimate for the
smaller vanilla+$N_{\rm eff}$+$^3f_\nu$ model, based on the MCI, is
$N_{\rm eff}= 3.2^{~4.5,~6.1}_{~3.0,~3.0}$ (All-Q+HST), while
for the larger vanilla+$N_{\rm eff}$+$^3f_\nu$+$\alpha_s$+$w$ model
we find $N_{\rm eff}=3.2^{~4.3,~5.8}_{~3.0,~3.0}$ using  the same data set.

\section{Conclusions}                          \label{sec:conclusions}

Motivated by several recent, seemingly conflicting inferences of the
cosmic radiation density (traditionally parameterised as the
effective number of neutrino species $N_{\rm eff}$) from
cosmological observations, we have re-examined the issue of
cosmological $N_{\rm eff}$ determination in great detail and
identified the reasons for the apparent discrepancies.

Using a minimal model with $N_{\rm eff}$ as the only nonstandard
parameter (i.e., vanilla+$N_{\rm eff}$), we find that the treatment
of scale-dependent biasing in the galaxy power spectrum data is
crucial to the derived value of $N_{\rm eff}$. The very high values
of $N_{\rm eff}$ found in Refs.~\cite{Seljak:2006bg,Mangano:2006ur}
for the WMAP+SDSS-DR2+SNIa data and the same model can be traced to
their treatment of the $Q_{\rm nl}$ parameter which quantifies the
level of bias correction. The prior on $Q_{\rm nl}$ imposed in these
studies, $Q_{\rm nl} = 10\pm5$, is significantly more
restrictive than the parameter space allowed by the WMAP+SDSS-DR2-Q data.
Because of a degeneracy between $N_{\rm eff}$ and $Q_{\rm nl}$,
such a restrictive prior cuts out much of the parameter region
that favours low values of $N_{\rm eff}$ and consequently
biases the inferred $N_{\rm eff}$ towards high values (figure~\ref{fig:QN}).
The use of restrictive priors on $Q_{\rm nl}$ is unjustified
when fitting nonstandard cosmologies, unless the priors have been
verified/supplemented by simulations or other means under {\it the
same} model assumptions.  In the absence of such information, the
best strategy is to use broad and uniform priors.

When the WMAP measurements are combined with any other single data
set (LSS, BAO, SNIa, or HST), we find that the inferred $N_{\rm
eff}$ is always compatible with the standard value $N_{\rm
eff}=3.046$ at 68\% C.L.\ or better, except for the combination
WMAP+Ly$\alpha$, which yields a high $N_{\rm eff}$ value in
disagreement with 3.046 at more than 95\%. The reason Ly$\alpha$
prefers a high $N_{\rm eff}$ originates in a well-known discrepancy
in the inferred small-scale fluctuation amplitude between the
SDSS-Ly$\alpha$  and the WMAP data. This can be understood from our
figure~\ref{fig:degeneracies}.

When all data sets (except Ly$\alpha$) are used in combination, we
find tighter bounds on $N_{\rm eff}$ that are, again, compatible
with $N_{\rm eff}=3.046$ at better than the 68\% level. This finding
is independent of whether we use galaxy power spectrum data only in
the strictly linear regime or also at higher values of $k$, as long
as scale-dependent bias is correctly taken into account. When
Ly$\alpha$ is added to the fit, the inferred $N_{\rm eff}$ is again
shifted to higher values because of Ly$\alpha$'s normalisation
discrepancy with WMAP. As discussed in section \ref{sec:lya} this
discrepancy is most likely due to unaccounted systematics in the
Ly$\alpha$ data. For this reason we quote a result without
Ly$\alpha$, $N_{\rm eff}=2.6^{~3.6,~4.8}_{~1.8,~1.1}$, as our best
current estimate of the constraints on $N_{\rm eff}$ in the minimal
vanilla+$N_{\rm eff}$ model from WMAP+LSS-Q+BAO+SNIa+HST.

Another very interesting point is that the statistical method used
to construct credible intervals can have a strong impact on
parameter inference when the posterior probability is non-Gaussian.
Using an inappropriate interval construction can sometimes lead to
incorrect inferences. This is especially true when fitting data sets
that are not very constraining and therefore contain strong
parameter degeneracies.   However, when all available data sets are
used in combination, they conspire to break each other's
degeneracies. The 1D posterior for $N_{\rm eff}$ in the minimal
model approaches the Gaussian limit, and all three interval
constructions used in our analysis, the Bayesian central and minimum
credible intervals, and the non-Bayesian concept of maximisation,
give almost identical results in this case.

New parameter degeneracies arise when more free parameters are
introduced in extended models. Even when the parameter inference is
performed with all data sets combined, there is still some, albeit
small, differences in the credible intervals obtained from the
different methods.   We have considered several different extended
models in the present work, all including nonzero neutrino masses as
a free parameter. While the formal constraints on $N_{\rm eff}$
differ slightly from model to model, we find again that $N_{\rm
eff}=3.046$ is always compatible with data at the 68\% C.L.\ or
better, as long as we exclude the Ly$\alpha$ data. Because of the
additional parameters the formal bounds on $N_{\rm eff}$ are
somewhat relaxed relative to those derived for the minimal model.
For our most general model (i.e., vanilla+$N_{\rm
eff}$+$f_\nu$+$\alpha_s$+$w$, with $N_{\rm eff}$ equally massive
neutrinos), we find $N_{\rm eff}=3.0^{~4.3,~6.1}_{~1.7,~0.8}$, based
on the minimum credible interval, using the data set
WMAP+LSS-Q+BAO+SNIa+HST.

We consider also the case in which the total radiation density is
split into three massive species and $N_{\rm eff}-3$ strictly
massless ones.  In this case we find almost identical upper bounds
on $N_{\rm eff}$ as in the previous case with $N_{\rm eff}$ massive
species (the lower bounds here are now always 3.0). Extra radiation
density corresponding to at least one extra neutrino degree of
freedom is allowed by all data sets at the 95\% level.  Thus,
cosmological observations are not yet at a precision level
sufficient to exclude very light sterile neutrinos, axions,
majorons, or similar particles that were in thermal equilibrium
after the QCD phase transition.  With future CMB and weak
gravitational lensing data this situation is set to change.  For
instance, with data from Planck and the future wide-field weak
lensing survey LSST, a sensitivity of $\sigma(N_{\rm eff}) \sim
0.07$ can be achieved~\cite{Hannestad:2006as}. Cosmology will then
become an even more powerful probe of particle physics beyond the
standard model.

\section*{Acknowledgements}

We thank An\v{z}e Slosar for useful suggestions and comments on
the manuscript.
We acknowledge use of computing resources from the Danish Center for
Scientific Computing (DCSC). In Garching and Munich, partial support
by the Deutsche Forschungsgemeinschaft under the grant TR~27
``Neutrinos and beyond'' and by the European Union under the ILIAS
project, contract No.~RII3-CT-2004-506222, is acknowledged.

\section*{References}


\begin{thebibliography}{99}

\bibitem{Tegmark:2006az}
  M.~Tegmark {\it et al.},
  ``Cosmological Constraints from the SDSS Luminous Red Galaxies,''
  Phys.\ Rev.\  D {\bf 74} (2006) 123507
  [arXiv:astro-ph/0608632].


\bibitem{nnu}
G.~Mangano, G.~Miele, S.~Pastor, T.~Pinto, O.~Pisanti and P.~D.~Serpico,
 ``Relic neutrino decoupling including flavour oscillations,''
  Nucl.\ Phys.\ B {\bf 729} (2005) 221 
  [arXiv:hep-ph/0506164].



\bibitem{bib:hannestad2003}
  S.~Hannestad,
  ``Neutrino masses and the number of neutrino species
  from WMAP and  2dFGRS,''
  JCAP {\bf 0305} (2003) 004
  [arXiv:astro-ph/0303076].




\bibitem{Crotty:2003th}
  P.~Crotty, J.~Lesgourgues and S.~Pastor,
  ``Measuring the cosmological background of relativistic particles with
  WMAP,''
  Phys.\ Rev.\  D {\bf 67} (2003) 123005 
  [arXiv:astro-ph/0302337].

\bibitem{Pierpaoli:2003kw}
  E.~Pierpaoli,
  ``Constraints on the cosmic neutrino background,''
  Mon.\ Not.\ Roy.\ Astron.\ Soc.\  {\bf 342} (2003) L63 
  [arXiv:astro-ph/0302465].

\bibitem{Barger:2003zg}
  V.~Barger, J.~P.~Kneller, H.~S.~Lee, D.~Marfatia and G.~Steigman,
  ``Effective number of neutrinos and baryon asymmetry from BBN and WMAP,''
  Phys.\ Lett.\  B {\bf 566} (2003) 8
  [arXiv:hep-ph/0305075].

\bibitem{Crotty:2004gm}
  P.~Crotty, J.~Lesgourgues and S.~Pastor,
  ``Current cosmological bounds on neutrino masses
  and relativistic relics,''
  Phys.\ Rev.\ D {\bf 69} (2004) 123007
  [arXiv:hep-ph/0402049].

\bibitem{Hannestad:2003ye}
  S.~Hannestad and G.~Raffelt,
  ``Cosmological mass limits on neutrinos, axions, and other light
  particles,''
  JCAP {\bf 0404} (2004) 008
  [arXiv:hep-ph/0312154].




\bibitem{Seljak:2006bg}
  U.~Seljak, A.~Slosar and P.~McDonald,
  ``Cosmological parameters from combining the Lyman-alpha forest
  with CMB, galaxy clustering and SN constraints,''
  JCAP {\bf 0610} (2006) 014
  [arXiv:astro-ph/0604335].

\bibitem{Hannestad:2006mi}
  S.~Hannestad and G.~G.~Raffelt,
  ``Neutrino masses and cosmic radiation density: Combined analysis,''
  JCAP {\bf 0611} (2006) 016
  [arXiv:astro-ph/0607101].

\bibitem{Cirelli:2006kt}
  M.~Cirelli and A.~Strumia,
  ``Cosmology of neutrinos and extra light particles after WMAP3,''
  JCAP {\bf 0612} (2006) 013
  [arXiv:astro-ph/0607086].

\bibitem{Ichikawa:2006vm}
  K.~Ichikawa, M.~Kawasaki and F.~Takahashi,
  ``Constraint on the effective number of neutrino species from
  the WMAP and SDSS LRG power spectra,''
JCAP {\bf 0705} (2007) 007
  [arXiv:astro-ph/0611784].

\bibitem{Mangano:2006ur}
  G.~Mangano, A.~Melchiorri, O.~Mena, G.~Miele and A.~Slosar,
  ``Present bounds on the relativistic energy density in the
  Universe from cosmological observables,''
  JCAP {\bf 0703} (2007) 006
  [arXiv:astro-ph/0612150].

\bibitem{Friedland:2007vv}
  A.~Friedland, K.~M.~Zurek and S.~Bashinsky,
  ``Constraining Models of Neutrino Mass and Neutrino Interactions with the
  Planck Satellite,''
  arXiv:0704.3271 [astro-ph].


\bibitem{Spergel:2006hy}
  D.~N.~Spergel {\it et al.}, ``Wilkinson Microwave Anisotropy Probe
  (WMAP) three year results: Implications for cosmology,''
Astrophys.\ J.\ Suppl.\ {\bf 170} (2007) 377
  [arXiv:astro-ph/0603449].


\bibitem{Hinshaw:2006ia}
  G.~Hinshaw {\it et al.},
  ``Three-year Wilkinson Microwave Anisotropy Probe (WMAP)
  observations: Temperature analysis,''
Astrophys.\ J.\ Suppl.\ {\bf 170} (2007) 288
  [arXiv:astro-ph/0603451].

\bibitem{Page:2006hz}
  L.~Page {\it et al.},
  ``Three year Wilkinson Microwave Anisotropy Probe (WMAP)
  observations: Polarization analysis,''
Astrophys.\ J.\ Suppl.\ {\bf 170} (2007) 335
  [arXiv:astro-ph/0603450].


\bibitem{Lewis:2002ah}
  A.~Lewis and S.~Bridle,
  ``Cosmological parameters from CMB and other data:
  A Monte-Carlo approach,''
  Phys.\ Rev.\ D {\bf 66} (2002) 103511 
  [arXiv:astro-ph/0205436]

\bibitem{cosmomc}
  A.~Lewis, Homepage, {\tt http://cosmologist.info}


\bibitem{mcdonald}
  P.~McDonald {\it et al.},
  ``The linear theory power spectrum from the Lyman-alpha
  forest in the Sloan Digital Sky Survey,''
  Astrophys.\ J.\  {\bf 635} (2005) 761
  [arXiv:astro-ph/0407377].


\bibitem{Tegmark:2003uf}
  M.~Tegmark {\it et al.} [SDSS Collaboration],
  ``The 3D power spectrum of galaxies from the SDSS,''
  Astrophys.\ J.\  {\bf 606} (2004) 702
  [arXiv:astro-ph/0310725].

\bibitem{Tegmark:2003ud}
  M.~Tegmark {\it et al.} [SDSS Collaboration],
  ``Cosmological parameters from SDSS and WMAP,''
  Phys.\ Rev.\  D {\bf 69} (2004) 103501
  [arXiv:astro-ph/0310723].


\bibitem{Hannestad:2005gj}
  S.~Hannestad,
  ``Neutrino masses and the dark energy equation of state:
  Relaxing the cosmological neutrino mass bound,''
  Phys.\ Rev.\ Lett.\  {\bf 95} (2005) 221301
  [arXiv:astro-ph/0505551].

\bibitem{lambda}
  Legacy Archive for Microwave Background Data Analysis
  (LAMBDA), \newline {\tt http://lambda.gsfc.nasa.gov}

\bibitem{Percival:2006gt}
  W.~J.~Percival {\it et al.},
  ``The shape of the SDSS DR5 galaxy power spectrum,''
Astrophys.\ J.\ {\bf 657} (2007) 645 
  [arXiv:astro-ph/0608636].

\bibitem{Cole:2005sx}
  S.~Cole {\it et al.} [2dFGRS Collaboration],
  ``The 2dF Galaxy Redshift Survey: Power-spectrum analysis
  of the final dataset and cosmological implications,''
  Mon.\ Not.\ Roy.\ Astron.\ Soc.\  {\bf 362} (2005) 505
  [arXiv:astro-ph/0501174].

\bibitem{Eisenstein2005}
  D.~J.~Eisenstein {\it et al.} [SDSS Collaboration],
  ``Detection of the baryon acoustic peak in the
  large-scale correlation function of SDSS
  luminous red galaxies,''
  Astrophys.\ J.\  {\bf 633} (2005) 560
  [arXiv:astro-ph/0501171];
  see also
  {\tt http://cmb.as.arizona.edu/$\sim$eisenste/acousticpeak}

\bibitem{halofit}
 R.~E.~Smith {\it et al.} [Virgo Consortium Collaboration],
  ``Stable clustering, the halo model and nonlinear cosmological
  power spectra,''
  Mon.\ Not.\ Roy.\ Astron.\ Soc.\ {\bf 341} (2003) 1311
  [arXiv:astro-ph/0207664].

\bibitem{Davis:2007na}
  T.~M.~Davis {\it et al.},
  ``Scrutinizing exotic cosmological models using ESSENCE
  supernova data combined with other cosmological probes,''
  arXiv:astro-ph/0701510.

\bibitem{Astier:2005qq}
  P.~Astier {\it et al.},
  ``The Supernova Legacy Survey: Measurement of $\Omega_M$,
  $\Omega_\Lambda$ and $w$ from the first year data set,''
  Astron.\ Astrophys.\  {\bf 447} (2006) 31
  [arXiv:astro-ph/0510447].

\bibitem{Wood-Vasey:2007jb}
  W.~M.~Wood-Vasey {\it et al.},
  ``Observational constraints on the nature of the dark energy:
  First cosmological results from the ESSENCE supernova survey,''
  arXiv:astro-ph/0701041.

\bibitem{Riess:2006fw}
  A.~G.~Riess {\it et al.},
  ``New Hubble Space Telescope discoveries of type Ia supernovae at $z > 1$:
  Narrowing constraints on the early behavior of dark energy,''
Astrophys.\ J.\ {\bf 659} (2007) 98
  [arXiv:astro-ph/0611572].

\bibitem{Freedman:2000cf}
  W.~L.~Freedman {\it et al.},
  ``Final results from the Hubble Space Telescope key project
  to measure the Hubble constant,''
  Astrophys.\ J.\ {\bf 553} (2001) 47
  [arXiv:astro-ph/0012376].


\bibitem{Viel:2005eg}
  M.~Viel, M.~G.~Haehnelt and V.~Springel,
  ``Testing the accuracy of the Hydro-PM approximation in
  numerical simulations of the Lyman-alpha forest,''
  Mon.\ Not.\ Roy.\ Astron.\ Soc.\  {\bf 367} (2006) 1655
  [arXiv:astro-ph/0504641].

\bibitem{Viel:2005ha}
  M.~Viel and M.~G.~Haehnelt,
  ``Cosmological and astrophysical parameters from the SDSS flux power spectrum
  and hydrodynamical simulations of the Lyman-alpha forest,''
  Mon.\ Not.\ Roy.\ Astron.\ Soc.\  {\bf 365} (2006) 231
  [arXiv:astro-ph/0508177].

\bibitem{viel2006}
  M.~Viel, M.~G.~Haehnelt and A.~Lewis,
  ``The Lyman-alpha forest and WMAP year three,''
  Mon.\ Not.\ Roy.\ Astron.\ Soc.\  {\bf 370} (2006) L51
  [arXiv:astro-ph/0604310].

\bibitem{Benson:1999mv}
  A.~J.~Benson, S.~Cole, C.~S.~Frenk, C.~M.~Baugh and C.~G.~Lacey,
  ``The nature of galaxy bias and clustering,''
  Mon.\ Not.\ Roy.\ Astron.\ Soc.\  {\bf 311} (2000) 793
  [arXiv:astro-ph/9903343].

\bibitem{Blanton:1999gd}
  M.~Blanton, R.~Cen, J.~P.~Ostriker, M.~A.~Strauss and M.~Tegmark,
  ``Time evolution of galaxy formation and bias in
  cosmological simulations,''
  Astrophys.\ J.\ {\bf 531} (2000) 1
  [arXiv:astro-ph/9903165].

\bibitem{Seljak:2000jg}
  U.~Seljak,
  ``Redshift space bias and beta from the halo model,''
  Mon.\ Not.\ Roy.\ Astron.\ Soc.\  {\bf 325} (2001) 1359
  [arXiv:astro-ph/0009016].

\bibitem{Smith:2006ne}
  R.~E.~Smith, R.~Scoccimarro and R.~K.~Sheth,
  ``The scale dependence of halo and galaxy bias:
  Effects in real space,''
  Phys.\ Rev.\  D {\bf 75} (2007) 063512
  [arXiv:astro-ph/0609547].

\bibitem{Cole:2006kn}
  S.~Cole, A.~G.~Sanchez and S.~Wilkins,
  ``The galaxy power spectrum: 2dFGRS-SDSS tension?,''
  arXiv:astro-ph/0611178.



\bibitem{Tegmark:2000db}
  M.~Tegmark and M.~Zaldarriaga,
  ``Current cosmological constraints from a 10 parameter CMB
  analysis,''
  Astrophys.\ J.\ {\bf 544} (2000) 30
  [arXiv:astro-ph/0002091].

\bibitem{Goobar:2006xz}
  A.~Goobar, S.~Hannestad, E.~M\"ortsell and H.~Tu,
  ``A new bound on the neutrino mass from the SDSS baryon acoustic peak,''
  JCAP {\bf 0606} (2006) 019
  [arXiv:astro-ph/0602155].

\bibitem{Hannestad:2006as}
  S.~Hannestad, H.~Tu and Y.~Y.~Y.~Wong,
  ``Measuring neutrino masses and dark energy with weak lensing tomography,''
  JCAP {\bf 0606} (2006) 025
  [arXiv:astro-ph/0603019].


\end{thebibliography}
\end{document}